# Employment, labor productivity and environmental sustainability: Firm-level evidence from transition economies

**Marjan Petreski**
University American College Skopje
marjan.petreski@uacs.edu.mk

**Stefan Tanevski**
University American College Skopje
stefan.tanevski@uacs.edu.mk

**Irena Stojmenovska**
University American College Skopje
irena.stojmenovska@uacs.edu.mk

## Abstract

This paper examines how investment in environmentally sustainable practices impacts employment and labor productivity growth of firms in transition economies. The study considers labor skill composition and geographical differences, shedding light on sustainability dynamics. The empirical analysis relies on the World Bank's Enterprise Survey 2019 for 24 transition economies, constructing an environmental sustainability index from various indicators through a Principal Components Analysis. To address endogeneity, a battery of fixed effects and instrumental variables are employed. Results reveal the relevance of environmental sustainability for both employment and labor productivity growth. However, the significance diminishes when addressing endogeneity comprehensively, alluding that any relation between environmentally sustainable practices and jobs growth is more complex and needs time to work. The decelerating job-creation effect of sustainability investments is however confirmed for the high-skill firms, while low-skill firms benefit from labor productivity gains spurred by such investment. Geographically, Central Europe sees more pronounced labor productivity impacts, possibly due to its higher development and sustainability-awareness levels as compared to Southeast Europe and the Commonwealth of Independent States.

**Keywords:** employment growth; labor productivity growth; environmental sustainability; transition economies

**JEL classification:** J24; O44

## 1. Introduction

The effect of environmental and other non-financial factors on firms' performance has recently captured academic and policy attention (Huang, 2021). Driven by the rising global awareness about the climate change, social responsibility and proper governance, the pressure onto firms has been increasing to incorporate principles of environmental, social and governance (ESG) sustainability into their work so as to deliver towards the society despite in an unparalleled way when compared to delivering to their shareholders (Gillan et al. 2010). However, the attainment of this objective requires investment on which the firm would expect certain return.

Literature on firms' value-profit maximization, on one hand, and societal contribution (going past the simple firm definition), on the other, persists in some form since the 1970s (Mackey, 2007). In our attempt to provide a comprehensive overview of ESG literature, we identified a strand of scholars who favor the traditional shareholder value maximization pillar (e.g. Friedman, 2007; Copeland et al. 1994) and those who believe that firms should be more than profit-oriented – being socially responsible even at a cost of forgoing present value of future cash flows (e.g. Freeman, 1984; Clarkson, 1995). Such early dissentions are what spurred later growth in academic research on this topic, resulting in a large body of literature attempting to resolve conflicting theories (Margolis and Walsh, 2003; Mitchell et al. 1997; Wood & Jones, 1995).

In this context, most academic research has focused on assessing ESG effects on a firm's financial performance and stock returns. Orlitzky et al. (2003) contributed significantly to this field with their extensive meta-analysis spanning three decades and incorporating 52 studies, totaling 33,878 observations. Their work established a robust and positive one-to-one correlation between corporate social performance and financial performance. Building on this foundation, Margolis et al. (2009) conducted a corroborative meta-analysis, incorporating data from 167 studies. Yet, evidence for developing countries is ambiguous. Research on multinationals in Latin America demonstrates negative effects of ESG investments: firms who required greater ESG investments often diverted resources from important projects (Duque-Grisales and Aguilera-Caracuel, 2019). This is conflicting with findings of Miralles-Quirós et al. (2018) who demonstrate a favorable correlation between ESG factors and the economic performance of publicly traded companies in Brazil, while Junius et al. (2020) find no significance of ESG for firm performance in developing Asia. Other prominent literature on the topic includes Margolis & Walsh (2003), Margolis et al. (2007), Goss and Roberts (2011), Deng et al. (2013).

In light of effects from ESG standards on stock prices, a more prominent mechanism for evaluation centers around investor sentiment. For instance, Cordeiro and Tewari (2015) find that firms that exhibit higher ESG rankings experience an increase in stock prices due to greater anticipation of increase in future cash flows which investors primarily attribute to environmentally-conscious consumers, employees, non-governmental organizations (NGOs), and regulatory authorities. Gao et al. (2022) find that ESG performance shields from stock price crash as it provides for more environmentally-conscious investors, increased competence among analysts, and better control of managerial governance. Yet, this is conditional on the level of environmental awareness in the firm's location (i.e. less pronounced in cities exhibiting reduced commitment to environmental legislation and higher degree of marketization). This paper diverts the academic attention from financial returns and stock prices to an area which remains under-researched: the potential relations between the ESG sustainability and labor-related outcomes. We are rather modest and focus on whether investment in environmental sustainability (the E from ESG) supports the growth of jobs and labor productivity, both in the context of firm skill endowment in transition economies. From that viewpoint, rather than shedding light on the financial return, the paper focuses on another important aspect of what may constitute a return from investment: quality employment growth. To our knowledge, no paper has had similar objective and as such represents novelty in its entirety. Moreover, the research brings novelties in some of its specifics aspects: includes the reliance on own-created index from a myriad of environmental sustainability aspects reported by firms; and the extensive treatment of environmental sustainability as endogenous with respect to jobs and labor productivity growth.

The relationship between employment and skills on one hand, and ESG practices has been scarcely examined in the current literature. For example, Barrymore and Sampson (2021) investigate if the investment in environmental and related factors in the firm results in better financial performance due to the mediating role of labor productivity. On one hand, firms with such practices may be more selective in hiring and better able to retain and engage (talented) employees. On the other hand, however, investment in environmental and non-financial performance may divert firm's attention and resources towards such outcomes and negatively impinge on revenue generation and hence employment growth. Yet, Barrymore and Sampson (2021) use revenue per employee, which is still a measure of financial performance rather than strictly measuring labor productivity.

A strand of (older) literature nevertheless builds a consensus on the positive impact of environmental standards on labor productivity. Delmas and Pekovic (2013) and Ambec and Lanoie (2008) find that this occurs as a consequence of employees' commitment to the firm as well as of increased training. Likewise, adoption of such standards is found to contribute to perceptions of firm integrity among employees and prospective job applicants (i.e. working in a 'greener' firm) which in turn affects employees' productivity (Rogovsky and Dunfee, 2002; Guiso et al. 2015; Barrymore and Sampson, 2021). Greater employee satisfaction coupled with increased commitment is exerted when a firm is perceived as a responsible corporate entity. Mitchell et al. (2001) and Peterson (2004) find that job embeddedness increases when there's a perceived alignment between employee's values and corporate culture, a type of satisfaction that contributes to enhancing employee retention and satisfaction.

In their attempt to assess effects of ESG on labor productivity, Delmas and Pekovic (2013) hypothesize that organizational changes associated with employee training and interpersonal contacts (i.e. adoption of ISO 14001), led to greater labor productivity by fostering knowledge transfer and innovative ideas. Barrymore and Sampson (2021) demonstrate statistically significant and positive relationship between ESG factors and labor productivity yet revealing sectoral heterogeneity. They find that variety in outcome is conditional on ESG components: e.g. in the mining sector, environmental elements of ESG exhibit greater variability in their association with labor productivity, while in the services sector, social elements show more pronounced fluctuations. Additionally, when accounting for the adoption of greenhouse gas emission among UK firms, they observe a significant drop in the labor productivity of high-emitting firms and an increase in the labor productivity of low-emitting ones, a pattern consistent when compared to a control group of European firms unaffected by the regulation.

Another relevant aspect for our study is that despite the topic of the environmental and other non-financial practices of firms have been around in the last two decades, its substantiation has been mainly noted in the advanced economies. Yet, globalization diffused through trade and MNCs, coupled with the weak institutional setting, requires that firms based in emerging and transition economies comprehensively address these challenges (Forcadell and Aracil, 2019). Hitherto, the literature on how ESG factors affect firm performance is mainly about advanced economies; recent examples include: Brogi and Lagasio (2019); Baraibar-Diez et al. (2019); Miralles-Quiroos et al. (2019). Nevertheless, the pertinent literature in developing economies is likewise emerging; prominent articles include Duque-Grisales and Aguilera-Caracuel (2019); Pollard et al. (2018); Shakil et al. (2019).

The paper is structured as follows: Section 2 presents the theoretical underpinnings of our analysis, i.e. nests the relevance of the environmental sustainability for the employment in a conceptual framework which produces the empirical framework. Section 3 presents the data availability and the necessities to create an environmental sustainability index. Section 4 discusses issues pertinent to the estimation of the empirical specification. Section 5 presents the results and offers a discussion. Section 6 concludes and provides some policy ramifications.

## 2. Theoretical and empirical framework

In our approach, we are more specific and focus on the employment growth and its skill composition, as well on the environmental sustainability only, and make the rather brave assumption that the potential effect of the environmental sustainability practices works for firm's employment performance and quality through affecting firm's efficiency in producing its products. The starting point is a production function whereby the current products in the first (current) and second (new) period are denoted as $Y_1$ and $Y_2$, respectively. Accordingly, the production in the first and second period are denoted as $Y_{11}$ and $Y_{12}$, respectively. The equations are given by

$$Y_{11i} = \theta_{11} F(L_{11i}, K_{11i}, M_{11i}) e^{\eta_i}$$

$$Y_{12i} = \theta_{12} F(L_{12i}, K_{12i}, M_{12i}) e^{\eta_i - u_i}$$

Whereby $F$ is homogeneous function in labor ($L$), capital ($K$) and intermediate goods ($M$); $\theta$ is a Hicks neutral technical change parameter; and $e^{\eta - u}$ is unobserved firm's productivity which is determined by firm's characteristics which are time-invariant ($\eta$) and productivity shocks ($u$). Under perfect competition in input markets, the cost function of a firm is

$$C_1(\omega_{11i}, Y_{11}) = F(L_{11i}, K_{11i}, M_{11i}) c(\omega_{11i}) = \frac{Y_{11i}}{\theta_{11} e^{\eta_i}} c(\omega_{11i})$$

$$C_2(\omega_{12i}, Y_{12}) = F(L_{12i}, K_{12i}, M_{12i}) c(\omega_{12i}) = \frac{Y_{12i}}{\theta_{12} e^{\eta_i - u_i}} c(\omega_{12i})$$

whereby $\omega_{11i}, \omega_{22i}$ are the input prices in the first and second period, respectively, and $c$ is the cost function to make the product.

Shephard's lemma states that the partial derivative of a firm's cost function with respect to a particular input (such as labor or capital) is equal to the marginal product of that input in the firm's production function. Applying Shephard's lemma (derivative of total cost over price of labor) we get the following equations for the labor:

$$L_{11i} = c_{\omega_L}(w_{11i}) \frac{Y_{11i}}{\theta_{11} e^{\eta_i}} \qquad L_{12i} = c_{\omega_L}(w_{12i}) \frac{Y_{12i}}{\theta_{12} e^{\eta_i - u_i}}$$

Now consider:

$$\frac{\Delta L_i}{L_i} = \frac{L_{12i} - L_{11i}}{L_{11i}} \approx ln\left(\frac{L_{12i}}{L_{11i}}\right)$$

If we substitute with the labor equations, we get to:

$$\frac{\Delta L_i}{L_i} = -(ln\theta_{12} - ln\theta_{11}) + (lnY_{12i} - lnY_{11i}) + u_i$$

The above can now be written as

$$l_i - y_{1i} = \alpha_0 + \alpha_1 d_i + u_i \quad y_{1i} = ln\left(\frac{Y_{12i}}{Y_{11i}}\right) \alpha_0 = -(ln\theta_{12} - ln\theta_{11}) \tag{1}$$

whereby $l$ is total employment growth, while $d$ captures the efficiency in the firm's processes, which we assume to have been modeled by the introduction of and investment in the environmental sustainability of the firm. The idea has a parallel in the innovation literature; e.g. Harrison et al. (2014) conceptualizes $d$ as the introduction of innovation in the current production process (i.e. not introducing new products), and hence we borrow and adjust the concept from there. By analogy, hence, the constant in (1) represents the average efficiency growth in the production for the firms who did not introduce / comply with the environmental-sustainability standards.

Therefore, our empirical model is as follows:

$$\Delta ep_i = \alpha_0 + \beta_1 \Delta y_i + \beta_2 env_i + \gamma_k \sum_{k=1}^{n} demo_{ki} + u_i \tag{2}$$

Whereby $\Delta ep_i$ is derived from: total employment growth and labor productivity growth; $\Delta y_i$ is the total sales growth; $env_i$ is a measure(s) of the environmental sustainability of the firm and $demo_{ki}$ is a vector of $n$ demographic characteristics of the firm. In this model, $env_i$ is the variable of our central interest.

Environmental sustainability processes in a firm may be defined as the various activities, practices, and strategies that organizations implement to manage their environmental impact and promote sustainability. Examples of such processes may be manifold and various. Firms need to comply with a myriad of environmental regulations and laws imposed by local, national, and international authorities. Compliance involves monitoring and ensuring adherence to standards related to emissions, waste disposal, water usage, and more (Ramanathan et al. 2017). Firms seek to optimize the use of natural resources, such as energy, water, and raw materials. This may involve implementing resource-efficient technologies, adopting recycling programs, and reducing resource wastage (Grejo, 2022). Firms work with their suppliers and logistics partners to reduce the environmental footprint of their supply chain. That includes sourcing sustainable materials, minimizing transportation emissions, and promoting responsible sourcing (Parashar, 2020; Ghosh and Shah, 2015). Firms often invest in energy-efficient technologies, conduct energy audits, and integrate renewable energy sources like solar and wind into their operations (Abdmouleh, 2015). Firms are increasingly focused on measuring and reducing their carbon emissions. This includes setting targets for emissions reduction, purchasing carbon offsets, and implementing transportation alternatives that reduce emissions (Qian and Schaltegger, 2017). Without going even forefather, the above imply that to capture environmental sustainability of the firm, we need to incorporate numerous aspects. At this stage, it is determined by the data we have on disposal, which we explain next.

### 3. Data and environment sustainability index

The empirical model (2) is estimated for a set of 24 transition economies, of the total of 29 recognized transition economies, hence we have near full coverage. We rely on the Enterprise Survey of the World Bank, for the year of 2019 and a total of 14,484 firms across the transition countries. It is only 2019 we work with, because it only encompasses the Green Economy Module, whereby we source information from. However, the survey has temporal questions in that it asks the current employment and sales, and the one of two years ago, hence growth variables could be constructed.

The Green Economy Module asks each firm a total of 50 questions on environmental standards, energy consumption, area regulation, area managerial practices etc., hence providing a rich dataset to capture environmental sustainability in a multi-facet manner. The latter is the key contribution of the study, particularly over the straightforward use of single self-reported ESG metrics that is popular in the current research. To contain our approach in manageable bounds, we select the following environmental sustainability aspects to be captured in the analysis (major part of which determined by the fact whether they refer to all firms in the sample) (**Table 1**; more information is provided in Annex 1):

Table 1 – Environment-related aspects available in the Enterprise Survey

| Label | Description |
|---|---|
| Environmental objectives | If the firm has strategic objectives that mention environmental or climate change issues |
| Environmental Manager/Department | If the firm has a manager responsible for environmental and climate change issues |
| Environmental Standard | Firm's customers require environmental certifications or adherence to certain environmental standards as a condition to do business with |
| Monitoring consumption | If and frequency of monitoring energy consumption |
| CO2 Emission | If the firm emits CO2 |
| Emission chain | If the firm monitored CO2 emission along its supply chain |
| Other pollutant | If the firm emitted pollutants other than CO2 |
| Measures adopted | Number of measures adopted |
| Renewable sources | If the firm used energy from its own renewable sources, such as power plants using solar, wind, hydro, biomass or geothermal energy |

*Source: Drafted by the authors from the Enterprise Survey (2019).*

While environmental sustainability is determined / shaped by all these aspects, it is rather a latent variable, roughly be thought of being inexistent when all of these aspects are inexistent or not applied in a single firm, up to a maximal value when all aspects are pursued by the firm, and for some of them with the maximal amount/frequency. All of the above measures are set so that when they take a value of 1 they work positively for the environmental sustainability; or they do so when growing. One common approach in such cases when the variable to be explained is unobserved, is to use Principal Component Analysis (PCA) to create an index from these independent variables. PCA is a technique that reduces the dimensionality of a dataset while retaining as much variation in the data as possible. It creates new variables, called principal components, which are linear combinations of the original variables. These principal components are orthogonal and uncorrelated, and they capture the most significant patterns of variation in the data (Greene, 2018).

Therefore, the environmental sustainability index we create is based on weights derived from the factor loadings on the first component of the PCA. The first component explains slightly over 30% of the common variance across the indicators, which may be considered sufficient, given the next component explains only about 12% and the explanatory power of the other components further rapidly declines. The loadings are presented in the following table (Table 2).

Table 2 – Factor loadings of the PCA

|  | Component 1 |
|---|---|
| Environmental objectives | 0.3873 |
| Environmental Manager/Department | 0.4030 |
| Environmental Standard | 0.3579 |
| Monitoring consumption | 0.2604 |
| CO2 Emission | 0.3533 |
| Emission chain | 0.3069 |
| Other pollutant | 0.2822 |
| Measures adopted | 0.3984 |
| Renewable sources | 0.1856 |

*Source: Authors' calculations.*

We observe that the factor loadings have the signs along our expectations, i.e. all of them are positive, signifying their association with the latent variable. The created index in this manner takes any value and implies that its values should be only understood in their relative meaning, i.e. when compared to each other. **Figure 1** presents the density distribution of the generated environmental sustainability index. Clearly, a large share of firms is concentrated in the far left part of the distribution, signifying none or low environmental sustainability awareness; essentially, these are the firms who answered negative or with the lowest grading on the indicators constituting the index. For example, about 15% of the firms answered 'no' to all questions which enter into the index. Then, the share of firms rapidly declines.

**Figure 1 – Environmental sustainability index**

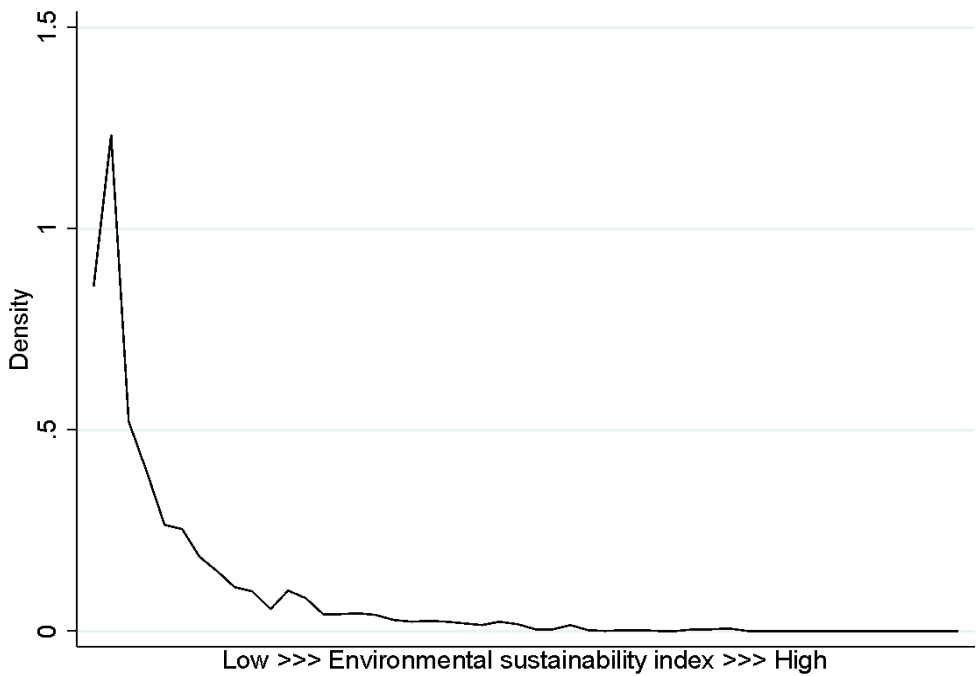

*Source: Authors' calculations.*

However, there could be important differences when firms are disaggregated by some observable characteristics. To illustrate any patterns, while also preserving the scarce space, we present the density distributions of the environmental sustainability index for firms divided by size and age. **Figure 2** (left) reveals that the environmental sustainability grows with size, as the density distribution becomes lower in the left part and higher in the right part as one goes from micro to large firm. Similarly, **Figure 2** (right) displays that old firms are more likely to take care of the environmental sustainability, without stark differences between young and mid-aged firms.

**Figure 2 – Environmental sustainability index, by firms' size (left) and age (right)**

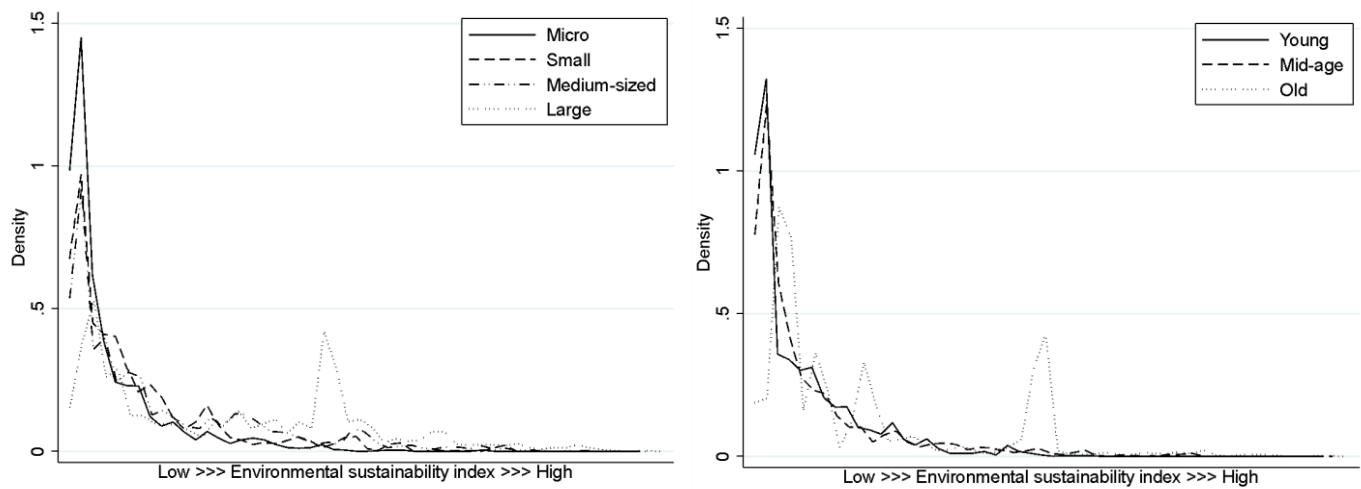

*Source: Authors' calculations.*

The other explanatory variables we use throughout the analysis are defined in a merely standard way and their details are provided in **Table A 1** in Annex 1. We control for the following demographic characteristics of the firm: ownership (domestic/foreign), female ownership, family ownership, age and size of the firm. It is critical to note that, as we self-generate the growth of employment and labor productivity, there appear cases of extremely low or extremely high growth, being determined by rapidly-declining or novel firms, respectively, most likely for factors unrelated to what we treat in this analysis. We therefore get rid of the outliers in both directions, through excluding the firms which fall in the lowest and highest 5% when ordered by the growth of employment or labor productivity.

The effect of the environmental sustainability onto employment and labor productivity growth, which considers the skill composition is estimated with a version of our model (2) for firms dominated by skilled, semi-skilled or unskilled workers, respectively, which is done by classifying each firm in one of the three categories of worker-skill composition depending on who dominates the structure. Also, countries are decomposed in three regional groups: Central Europe (CEE), Southeast Europe (SEE) and the Commonwealth of Independent States (CIS), potentially reflecting the different speeds of transition and stages of development they went through/are in.

## 4. Estimation considerations

The main estimation concern of the equation is the key variable potentially suffering endogeneity bias, stemming from the fact that overall productivity is omitted in equation (2), while it could be an important predictor of the introduction of and upgrade in the environmental sustainability aspects in the firm. To address this problem, we follow Elejalde et al. (2015) and decompose productivity in two unobserved components: firm's attributes that are mainly time invariant (like managerial skills or corporate culture) and productivity shocks. Our equation is set as a growth equation and hence the invariant part of productivity is, at least, attenuated in the error term. However, the correlation between productivity shocks and environmental sustainability may be harder to remove. To approach the issue in a compelling manner, it may be wise to recognize that both are related with the business cycle and/or with the functioning of the entire system which may be supportive or not to firms introducing environmental sustainability standards in their operations. Taking the recent surge in energy prices as a productivity shock, we may strongly assume that it also affects the intention of the firm to invest in environmental sustainability: it may either divert its attention from them to sustain the immediate surging operation costs, or may increase the focus on them as vehicle to increase operational efficiency over the medium run.

We will approach the problem in three ways. The first is through introducing industry dummies, as in Elejalde et al. (2015), because the industry dummies may be suitable in capturing the business cycle effect. The second is through introducing country dummies, because they will capture the government policies attributable to environmental sustainability. The third approach is through instrumenting the self-generated environmental sustainability index with an exogenous variation. Certainly, the latter is always challenging in empirical work and under the spotlight of academic reviewers, as exogenous variation may be hard to justify unless some statistical tests are assisting the argument. However, the conceptualization of the instrument is more important. A good instrument is the one that is correlated with the environmental sustainability index but is not correlated with the shocks onto employment or labor productivity growth. In our case, we are bound to seek exogenous instrument within the current survey, being a micro-survey of firms, since any instrument outside the survey would be at the industry or country level, hence captured by the respective fixed effects.

The Enterprise Survey contains the following questions: "Over the last three years, did this establishment experience monetary losses due to extreme weather events (such as storms, floods, droughts, or landslides)?" and "Over the last three years, did this establishment experience monetary losses due to pollution not generated by this establishment (that is, independent of this establishment's activity)?", which we deem suitable instruments. Firstly, environmental sustainability is about minimizing negative impacts on the environment. Occurrence of extreme weather events or exposure to pollution generated outside the firm can change the firm's stance towards the environment and make it more responsive to and/or aware of environmental sustainability. Hence, the relationship between environmental sustainability of the firm and extreme weather and external pollution is rather clear. However, secondly, extreme weather events or external pollution can have substantial financial consequences for firms. These events can damage infrastructure, disrupt operations, and lead to direct monetary losses, including repair costs and revenue reductions. This in turn may impinge onto employment growth and growth of labor productivity. Firms that are less environmentally sustainable may be more vulnerable to such losses due to inadequate environmental risk management (Uitto and Shaw, 2016). Thirdly, extreme weather events are typically random events that are not directly influenced by labor-related factors, while external pollution is determined by other players on the market hence being an exogenous event especially for small firms operating on a competitive market. Therefore, it is very likely that the extreme weather events and the external pollution affect growth of employment and labor productivity through the manner in which the firm handles environmental sustainability factors, and not directly, making them suitable instruments.

We need a further layer of caution, nevertheless. The instruments used in the IV estimation are indicators of the firm bearing of the monetary losses incurred by extreme weather and/or external pollution, not of the occurrence of the exogenous events solely. So, the identification strategy relies on bearing of such monetary losses being exogenous once we control for industry, country, age, size, and time-invariant productivity. We believe this is a still valid assumption because more productive and larger firms may be more capable to sustain such monetary losses. Given that we control for invariant productivity and size, these effects are taken into account. In addition, it seems less likely that firms decide to bear the monetary loss or not because of productivity shocks that could be temporary. In the following sections, we provide also some statistical tests about instruments' exogeneity.

## 5. Results and discussion

The results are presented in Table 3, for the two dependent variables and in four estimable forms: basic OLS estimation with no controls (columns 1 and 6), demographic variables added (columns 2 and 7), industry fixed effects added (columns 3 and 8), country fixed effects added (columns 4 and 9) and environmental sustainability instrumented (columns 5 and 10). Results reveal that the sales growth has consistently positive and significant relation with the employment and labor productivity growth: when sales grow by 1%, jobs grows by 0.16%, while labor productivity by 0.62%. For the IV estimations, towards the bottom of Table 3, the respective tests suggest that instruments are valid.

Environmental sustainability is relevant for both the employment growth and for the labor productivity growth, which largely corroborates existing empirical results (e.g. Kunapatarawong and Martinez-Ros, 2016; Delmas and Pekovic, 2013). Environmental sustainability index is, however, negatively related with the jobs growth, which may signify that investments in sustainability involved the adoption of automated or energy-efficient technologies that likely led to labor-saving effects through curbing employment growth not necessarily through job losses. Yet, the significance is lost in the IV estimation, which is not a result to be ignored, as potential endogeneity is treated in the fullest manner. First, we noticed in the earlier figures that most of the transition-country firms are concentrated towards the bottom part of the environmental sustainability index, suggesting they committed no or low investment in environmental sustainability. Smaller investments may not have a substantial impact on employment growth compared to larger-scale initiatives. Secondly, investments in environmental sustainability often yield long-term benefits, which may not manifest immediately in terms of increased employment growth. This is particularly relevant in our case, since we operate with only one year of the Enterprise Survey, which is the maximum data we have on disposal.

Thirdly and particularly relevant when operating in a transition economy context, firms may prioritize cost control and profitability over environmental sustainability. They may be hesitant to allocate significant resources to sustainability efforts if the expected return on investment is uncertain or takes a long time to realize. This has been corroborated in the empirical literature: Epstein et al. (2015) found that managers choose profitability over sustainability whenever they are in conflict.

A fourth plausible reason for the insignificance of the environmental sustainability investment for employment growth is that such investments may focus on improving labor productivity rather than increasing the number of employees. This is a more plausible explanation in our context, because the composition of the index is not based on variables capturing (large) investment in energy-efficient machines and processes per se, but rather considers more qualitative aspects of the sustainability, which may be more relevant for the labor productivity. Greater efficiency and reduced waste likely led to improved output without a corresponding increase in the workforce, which is documented through the positive and consistently significant coefficients on the environmental sustainability index in columns (6)-(10). The coefficient is of similar magnitude across specifications, confirming the robust labor-productivity-enhancing effect of the investment in environmental sustainability, despite its significance only at the 10% should be noted. The result is in line with e.g. Delmas and Pekovic (2013).

Table 3 – Results

| | Dependent variable: Employment growth | | | | | Dependent variable: Labor productivity growth | | | | |
|---|---|---|---|---|---|---|---|---|---|---|
| | (1) | (2) | (3) | (4) | (5) | (6) | (7) | (8) | (9) | (10) |
| Sales growth | 0.162*** | 0.149*** | 0.151*** | 0.159*** | 0.161*** | 0.598*** | 0.619*** | 0.612*** | 0.614*** | 0.618*** |
| | (0.048) | (0.047) | (0.049) | (0.049) | (0.049) | (0.058) | (0.056) | (0.059) | (0.060) | (0.060) |
| Environmental sustainability | -0.792* | -0.830* | -0.767* | -0.797* | 1.071 | 0.669* | 0.734* | 0.811* | 0.786* | 0.874* |
| | (0.418) | (0.448) | (0.454) | (0.451) | (2.559) | (0.389) | (0.413) | (0.424) | (0.430) | (0.580) |
| Demographic variables | No | Yes | Yes | Yes | Yes | No | Yes | Yes | Yes | Yes |
| Industry fixed effects | No | No | Yes | Yes | Yes | No | No | Yes | Yes | Yes |
| Country fixed effects | No | No | No | Yes | Yes | No | No | No | Yes | Yes |
| Instrument | No | No | No | No | Yes | No | No | No | No | Yes |
| R-squared | 0.04 | 0.05 | 0.06 | 0.07 | 0.04 | 0.27 | 0.28 | 0.29 | 0.30 | 0.30 |
| Underidentification test (Kleibergen-Paap rk LM statistic), P-val | | | | | 0.0563 | | | | | 0.0339 |
| Weak identification test (Kleibergen-Paap rk Wald F statistic) | | | | | 4.147 | | | | | 5.23 |
| Hansen J statistic (overidentification test of all instruments, P-val | | | | | 0.928 | | | | | 0.643 |

Source: Authors' calculations.

*, ** and *** refer to statistical significance at the 10, 5 and 1% level, respectively. Standard errors provided in parentheses. Standard errors robust to heteroscedasticity.

Results vary across skill-based groups, presented in Table 4. Recall, we classify each firm in high-, medium- and low-skill depending on which level of skills prevails in the employed labor force. It is only in the high-skill firms where the environmental sustainability exerts negative effect for the pace of job creation. This is plausible, since it is more likely that firms based on high-skilled workers to be the frontrunners of the investment in environment sustainability, and who are likely tilted towards capital-intensive sectors (like electrical equipment within manufacturing, or IT within services), which implies that their labor needs are different than the ones in the labor-intensive sectors, and further attenuated by the investment in energy-efficient technology and practices.

The opposite is true when the labor productivity growth is observed. Results suggest that investment in environmental sustainability practices may lead to significant productivity gains only in the low-skill firms. This implies that sectors which are mainly labor-intensive like textiles from manufacturing and trade and transport from services may positively contribute towards labor productivity through investment in environmentally sustainable practices. When these industries invest in environmentally sustainable practices, they frequently focus on improving resource efficiency. This includes reducing energy consumption, optimizing material usage, and minimizing waste. These efficiency gains can lead to higher labor productivity as employees work in a more streamlined and less wasteful environment (Chung et al. 2013). Additionally, these investments can lead to technological advancements, improved worker well-being, and reduced downtime, all of which contribute to higher labor productivity.

The geographical disaggregation (Table 5) suggests that it is only labor productivity growth in Central Europe for which the investment in environmental sustainability practices matter. This may imply that the effect of such investment for labor productivity starts in later transition – as Central Europe is most advanced among the three regional groupings – and from low-skill sectors. For the higher-skill sectors in these advanced transitioners, it may be that labor productivity and jobs growth are determined predominantly by other factors and the reconciliation of the environmentally sustainable practices with the firm growth model is irrelevant for the labor outcomes. For the other transitioners – Southeast Europe and the Commonwealth of Independent States, it may simply be that that investment in environmentally sustainable practices is in its nascence, so that tangible results are not yet observed.

Table 4 – Results, by skill level

| | Dependent variable: Employment growth | | | | | | Dependent variable: Labor productivity growth | | | | | |
|---|---|---|---|---|---|---|---|---|---|---|---|---|
| | High-skill | | Medium-skill | | Low-skill | | High-skill | | Medium-skill | | Low-skill | |
| | Non-IV | IV | Non-IV | IV | Non-IV | IV | Non-IV | IV | Non-IV | IV | Non-IV | IV |
| | (1) | (2) | (3) | (4) | (5) | (6) | (7) | (8) | (9) | (10) | (11) | (12) |
| Sales growth | 0.191*** | 0.187*** | 0.0921* | 0.0813 | 0.11 | 0.101 | 0.528*** | 0.552*** | 0.569*** | 0.569*** | 0.626*** | 0.624*** |
| | (0.048) | (0.059) | (0.050) | (0.064) | (0.069) | (0.075) | (0.059) | (0.070) | (0.123) | (0.128) | (0.073) | (0.074) |
| Environmental sustainability | -0.0714** | -0.316** | -0.00532 | -0.144 | -0.0708* | 0.0327 | -0.0286 | 0.348 | 0.00896 | 0.0213 | 0.0754* | 0.194* |
| | (0.033) | (0.139) | (0.052) | (0.359) | (0.042) | (0.098) | (0.039) | (0.217) | (0.054) | (0.378) | (0.039) | (0.031) |
| | | | | | | | | | | | | |
| R-squared | 0.223 | -0.295 | 0.094 | 0.027 | 0.167 | 0.11 | 0.314 | 0.025 | 0.321 | 0.321 | 0.434 | 0.396 |
| Underidentification test (Kleibergen-Paap rk LM statistic), P-val | | 0.000667 | | 0.313 | | 0.0079 | | 0.00405 | | 0.354 | | 0.0242 |
| Weak identification test (Kleibergen-Paap rk Wald F statistic) | | 12.21 | | 1.171 | | 10.5 | | 6.69 | | 1.035 | | 5.665 |
| Hansen J statistic (overidentification test of all instruments, P-val | | . | | 0.245 | | 0.0722 | | . | | 0.481 | | 0.567 |

Source: Authors' calculations.
*, ** and *** refer to statistical significance at the 10, 5 and 1% level, respectively. Standard errors provided in parentheses. Standard errors robust to heteroscedasticity. Demographic controls are included but not shown to preserve space.

Table 5 – Results, by geographic group

| | Dependent variable: Employment growth | | | | | | Dependent variable: Labor productivity growth | | | | | |
|---|---|---|---|---|---|---|---|---|---|---|---|---|
| | CEE | | SEE | | CIS | | CEE | | SEE | | CIS | |
| | Non-IV | IV | Non-IV | IV | Non-IV | IV | Non-IV | IV | Non-IV | IV | Non-IV | IV |
| | (1) | (2) | (3) | (4) | (5) | (6) | (7) | (8) | (9) | (10) | (11) | (12) |
| Sales growth | 0.182** | 0.190** | 0.252*** | 0.270*** | 0.0961** | 0.118** | 0.626*** | 0.640*** | 0.521*** | 0.514*** | 0.629*** | 0.626*** |
| | (0.073) | (0.075) | (0.040) | (0.048) | (0.047) | (0.058) | (0.086) | (0.092) | (0.057) | (0.056) | (0.058) | (0.058) |
| Environmental sustainability | -0.0334 | 0.0334 | -0.0241 | -0.313 | 0.000191 | 0.337 | 0.0613** | 0.141** | 0.00046 | 0.209 | 0.000676 | -0.0651 |
| | (0.031) | (0.111) | (0.028) | (0.207) | (0.027) | (0.347) | (0.031) | (0.075) | (0.034) | (0.288) | (0.029) | (0.269) |
| R-squared | 0.08 | 0.061 | 0.153 | -0.085 | 0.096 | -0.361 | 0.286 | 0.27 | 0.271 | 0.191 | 0.361 | 0.352 |
| Underidentification test (Kleibergen-Paap rk LM statistic), P-val | | 0.106 | | 7.01E-03 | | 0.00407 | | 0.068 | | 0.0204 | | 0.00605 |
| Weak identification test (Kleibergen-Paap rk Wald F statistic) | | 4.892 | | 6.884 | | 5.793 | | 8.386 | | 4.765 | | 5.068 |
| Hansen J statistic (overidentification test of all instruments, P-val | | 0.831 | | 0.754 | | 0.414 | | 0.766 | | 0.165 | | 0.872 |

Source: Authors' calculations.
*, ** and *** refer to statistical significance at the 10, 5 and 1% level, respectively. Standard errors provided in parentheses. Standard errors robust to heteroscedasticity. Demographic controls are included but not shown to preserve space.

## 6. Conclusion

The objective of the paper is to understand if investment in environmentally sustainable practices among firms in transition economies has a role to play for the employment and labor productivity growth. It sheds light onto the relationship in the context of firms' labor skill composition, as well among different geographical groupings. It has been hard to properly nest the paper within the current literature, which is emerging yet scant. Most of the studies examine the role of environmentally sustainable practices for financial aspects of firms, while when job generation is examined, this is usually in the context of the green transition determining closure of energy producers based on fossil fuels towards investment in renewable energy producers. The latter, however, is fully out of the scope of our paper. Instead, we focused on standard firms, most of whom nested in industry and services, to check if their job creation and labor productivity growth may be accelerated or not by investment in environmental sustainability.

Our theoretical approach is based in the innovation literature, recognizing investment into environmentally sustainable practices as innovative processes. Then, employment and labor productivity growth are determined by the growth of sales and efficiency in the firm's processes, which we assume to have been modelled by the introduction of and investment in the environmental sustainability practices. The empirical approach has been partially shaped by the availability of data. Namely, we work with the Enterprise Survey of the World Bank, which in its 2019 introduced a Green Economy Module, which comprehends about 50 questions on various aspects of environmental standards, energy-saving investment and practices, energy consumption measurement and practices, and so on. This implied that we did not have one single indicator to measure the extent to which a firm is environmentally sustainable, but rather a battery of indicators. The endeavor, hence, has been to combine finite number of indicators in a single index. Assuming that environmental sustainability at the firm level is a latent process which extends over a continuum between two extremes, we applied a Principal Components Analysis to generate the index.

When such self-generated index is next plugged in the regression to explain employment and labor productivity growth, the methodological challenge is to understand its exogeneity in the process. To address this problem, we decompose productivity in two unobserved components: firm's attributes that are mainly time invariant and productivity shocks. As the latter and the environmental sustainability are likely related with the business cycle and/or the functioning of the entire system who may be supportive or not to firms introducing environmental aspects in their operations, we made use of 1) industry fixed effects, 2) country fixed effects, and 3) instrumental variables. We rely on the indicator of if monetary losses were incurred due to extreme weather events and due to other pollutants, as exogenous variations, which we argued for and provided statistical tests.

Results suggest that environmental sustainability is relevant for both the employment and for the labor productivity growth. However, the significance for the employment growth is lost when endogeneity is treated in the fullest manner. This may be due to still low or no investment in environmental sustainability, or even when doing so due to the time it needs to work for acceleration of the job creation. Yet, it could be also due to such investments focusing on improving labor productivity rather than increasing the number of employees. Indeed, higher environmental efficiency likely led to improved output without a corresponding increase in the workforce, which is revealed by the positive role that environmental sustainability is found to play in transition economies.

Observed by skill level, results suggest that it is only in the high-skill firms where the environmental sustainability exerts negative effect for the pace of job creation. This implies that firms which are likely tilted towards capital-intensive sectors have different labor needs than those in the labor-intensive sectors, and further attenuated by the investment in energy-efficient technology and practices. The opposite is true when the labor productivity growth is observed. Results suggest that investment in environmental sustainability practices may lead to significant productivity gains only in the low-skill firms. This implies that sectors which are mainly labor-intensive may positively contribute towards labor productivity through investment in environmentally sustainable practices. These investments can lead to energy efficiency, technological advancements, improved worker well-being, and reduced downtime, all of which contribute to higher labor productivity.

Observed geographically, investment in environmental sustainability practices matters only for the labor productivity growth in Central Europe, which may be related to the development level, i.e. that such effect starts working only in later transition and from low-skill sectors. For the other transitioners – Southeast Europe and the Commonwealth of Independent States, it may simply be that that investment in environmentally sustainable practices is in its nascence, so that tangible results are not yet observed.

Overall, the findings within the research underscore the importance of a nuanced and context-specific approach to environmental sustainability policies, recognizing that the impact on employment and labor productivity growth can vary depending on factors such as industry type, skill levels, and regional development stages. Policymakers shall consider these nuances when designing and implementing strategies to encourage sustainability investments and maximize their positive effects on both environmental and labor-related outcomes.

# Annex 1

## Table A 1 – Variables' definitions

| Variable | Question(s) from the survey | Rationale (if any) |
|---|---|---|
| **Dependent variables** | | |
| Employment growth | L1. At the end of fiscal year [Insert last complete fiscal year], how many permanent, full-time individuals worked in this establishment? Please include all employees and managers / L2. Looking back, at the end of fiscal year [Insert last complete fiscal year minus two], how many permanent, full-time individuals worked in this establishment? Please include all employees and managers | Employment growth is calculated over the previous two years, as per the provisions of the questionnaire. |
| Labor productivity growth | D2. In fiscal year [Insert last complete fiscal year], what were this establishment's total annual sales for all products and services? / L.1 At the end of fiscal year [Insert last complete fiscal year], how many permanent, full-time individuals worked in this establishment? Please include all employees and managers. (Respectively, N3 / L2) | Productivity per worker, computed as the total annual sales over the number of workers. Productivity growth is calculated over the previous two years, as per the provisions of the questionnaire. |
| **Explanatory variables** | | |
| Sales growth | D2. In fiscal year [Insert last complete fiscal year], what were this establishment's total annual sales for all products and services? / N3. Looking back to fiscal year [Insert last complete fiscal year minus two], what were total annual sales for this establishment? | Sales growth is calculated over the previous two years, as per the provisions of the questionnaire. |
| **Environmental sustainability index** | | |
| Environmental objectives | BMGA.1 In fiscal year [Insert last complete fiscal year], did this firm have strategic objectives that mention environmental or climate change issues? | Indicates general awareness of Environmental Standards and if such is a firm objective. |
| Environmental Manager/Department | BMGA.2 In fiscal year [Insert last complete fiscal year], did this establishment have a manager responsible for environmental and climate change issues? | Indicates a higher level of commitment compared to having general objectives that may lack prioritization and action. |
| Environmental Standard | BMGA.4 In fiscal year [Insert last complete fiscal year], did any of the establishment's customers require environmental certifications or adherence to certain environmental standards as a condition to do business with this establishment? | This implies that customers are considering the environmental impact of their business partners and may prefer to engage with firms that have strong ESG practices. |
| Monitoring consumption | BMGC.2 Over the last three years, how often did this establishment monitor its energy consumption? | Regular monitoring indicates that the firm is actively tracking its environmental impact and is likely to be more conscious of energy efficiency and sustainability. |
| CO2 Emission | BMGC.7 Over the last three years, did this establishment emit CO2? | Fundamental indicator of the establishment's environmental awareness and commitment to tracking its carbon emissions |
| Emission chain | Over the last three years, did this establishment monitor CO2 emission along its supply chain? | It accounts for emissions associated with the production of raw materials, transportation, manufacturing, and distribution, a more comprehensive view. |
| Other pollutant | Over the last three years, did this establishment emit pollutants other than CO2? (Such as other types of air pollutants, soil or land pollutants, and water pollutants) | Demonstrates greater devotion and environmental awareness beyond tracking CO2 emissions. |
| Measures adopted | BMGC.23 Over the last three years, did this establishment adopt any of the following measures? | Counting the number of measures adopted. Adoption of more measures implies greater devotion and environmental awareness. |
| Renewable sources | BMGE.5 In fiscal year [Insert last complete fiscal year], did this establishment use energy from its own renewable sources, such as power plants using solar, wind, hydro, biomass or geothermal energy? | Indicative of firms with high environmental awareness. Usage of renewable resources for operations would imply an environmentally advanced stage. |
| **Demographic variables** | | |
| Domestic Foreign State | B.2 What percentage of this firm is owned by each of the following: domestic, foreign, state-owned. | The share of each category of ownership exceeds 50%. |
| Female | B.4 Amongst the owners of the firm, are there any females? | The share of females among owners is larger than 50%. |

| Family | BMB.1 What percentage of the firm is owned by the same family? | The share of family members among owners is larger than 50%. |
|---|---|---|
| Firm Age | B.5 In what year did this establishment begin operations? Classified in young (<10 years), mid-aged (10-50) and old firms (> 50) | Firm age. (i.e., older firms may be characterized with more experience, better market positioning) |
| Firm Size | L.1 At the end of fiscal year [Insert last complete fiscal year], how many permanent, full-time individuals worked in this establishment? Please include all employees and managers Classified in micro (<10 employed), small (10-50) and medium-sized (50-250) and large firms (> 250) | Firm size. Large firms may have different patterns in employment/labor productivity growth than small firms. |
| **Instrumental variables** | | |
| Extreme weather | BMGb.1 Over the last three years, did this establishment experience monetary losses due to extreme weather events (such as storms, floods, droughts, or landslides)? | External events not associated with the specific firm, yet correlated with the environmental sustainability practices, in that that could spur its adoption, or that such practices' absence may rather increase environmental sustainability. Moreover, uncorrelated with the shocks onto jobs and labor productivity growth. |
| Other pollutants | BMGb.2 Over the last three years, did this establishment experience monetary losses due to pollution not generated by this establishment (that is, independent of this establishment's activity)? | |
| **Division based on skill intensity** | | |
| High Skilled | L.4a1 In highly skilled jobs, that is professionals whose tasks require extensive theoretical and technical knowledge | Adoption of environmental standards and practices may be skill-biased in that new standards and innovations require skilled labor. This would help identify if such trend is present among firms in transition economies. |
| Medium Skilled | L.4a2 In semi-skilled jobs, that is technicians whose tasks require some level of mechanical or technical knowledge | |
| Low Skilled | L.4ab In unskilled jobs, those whose tasks involve no specialized knowledge | |
| **Division based on geography** | | |
| CEE | Central Europe: Croatia, Czechia, Estonia, Hungary, Latvia, Lithuania, Poland, Romania, Slovenia | Adoption of environmental standards and practices may be related to geographical grouping as they exert different paces of development. |
| SEE | Southeast Europe: Albania, Bosnia and Herzegovina, Kosovo, Montenegro, North Macedonia, Serbia | |

| CIS | Commonwealth of Independent States: Azerbaijan, Georgia, Kazakhstan, Kyrgyz Republic, Moldova, Russia, Tajikistan, Ukraine, Uzbekistan | |

Table A 2 – Descriptive statistics

| Variable | Obs | Mean | Std. Dev. | Min | Max |
| --- | --- | --- | --- | --- | --- |
| Employment growth | 10,694 | 6.58 | 15.03 | (24.68) | 50.00 |
| Labor productivity growth | 9,086 | 8.58 | 21.38 | (33.33) | 66.67 |
| Sales growth | 10,097 | 11.72 | 18.75 | (24.22) | 66.67 |
| Environmental sustainability index | 14,484 | 0.00 | 1.66 | (1.50) | 9.02 |
| Environmental objectives | 14,484 | 0.17 | 0.38 | 0 | 1 |
| Environmental Manager/Department | 14,484 | 0.12 | 0.32 | 0 | 1 |
| Environmental Standard | 14,484 | 0.15 | 0.35 | 0 | 1 |
| Monitoring consumption | 14,484 | 2.25 | 1.94 | 0 | 8 |
| CO2 Emission | 14,484 | 0.22 | 0.56 | 0 | 2 |
| Emission chain | 14,484 | 0.04 | 0.20 | 0 | 1 |
| Other pollutant | 14,484 | 0.08 | 0.37 | 0 | 2 |
| Measures adopted | 14,484 | 2.97 | 2.77 | 0 | 10 |
| Renewable sources | 14,484 | 0.04 | 0.20 | 0 | 1 |
| Domestic ownership | 14,484 | 0.89 | 0.31 | 0 | 1 |
| Foreign ownership | 14,484 | 0.07 | 0.25 | 0 | 1 |
| State ownership | 14,484 | 0.01 | 0.08 | 0 | 1 |
| Female owner | 14,484 | 0.33 | 0.47 | 0 | 1 |
| Family owner | 14,484 | 0.42 | 0.49 | 0 | 1 |
| Young firm | 14,484 | 0.30 | 0.46 | 0 | 1 |
| Mid-aged firm | 14,484 | 0.67 | 0.47 | 0 | 1 |
| Old firm | 14,484 | 0.03 | 0.17 | 0 | 1 |
| Micro firm | 14,484 | 0.30 | 0.46 | 0 | 1 |
| Small firm | 14,484 | 0.41 | 0.49 | 0 | 1 |
| Medium-sized firm | 14,484 | 0.23 | 0.42 | 0 | 1 |
| Large firm | 14,484 | 0.07 | 0.25 | 0 | 1 |
| Losses due to pollution by others | 14,484 | 0.02 | 0.15 | 0 | 1 |
| Losses due to extreme weather | 14,484 | 0.08 | 0.28 | 0 | 1 |